\documentclass[prl,letterpaper,twocolumn,showpacs,superscriptaddress,amsmath,amsfonts,amssymb,preprintnumbers,citeautoscript]{revtex4}
\pdfoutput=1 
\usepackage[T1]{fontenc}\usepackage[latin1]{inputenc}
\usepackage{dcolumn,graphicx,color,booktabs,microtype,afterpage}
\usepackage[autoplay]{animate}
\usepackage[charter,greekuppercase=italicized]{mathdesign}\usepackage{sidecap}
\usepackage[shortlabels]{enumitem}
\renewcommand{\figurename}{Fig.}
\renewcommand{\tablename}{Table}
\makeatletter\renewcommand{\fnum@figure}[1]{\figurename~\thefigure~(color online).}\makeatother
\makeatletter\renewcommand{\fnum@table}[1]{\tablename~\thetable.}\makeatother

\newcount\hh \newcount\mm
\hh=\time \divide\hh by 60
\mm=\hh \multiply\mm by 60 \mm=-\mm
\advance\mm by \time
\def\now{\number\hh:\ifnum\mm<10{}0\fi\number\mm}

\usepackage[colorlinks,plainpages=false,linkcolor=blue,urlcolor=blue,citecolor=blue,pdfpagemode=UseNone,pdfstartview=FitBH]{hyperref}

\makeatletter
\newcommand{\bibstyle@supplement}{\bibpunct[, ]{[S}{]}{;}{n}{,}{, S\hspace{-2.1pt}}%
\gdef\bibnumfmt##1{[S##1]}}
\makeatother

\begin{document}

\makeatletter\renewcommand{\ps@plain}{%
\def\@evenhead{\hfill\itshape\rightmark}%
\def\@oddhead{\itshape\leftmark\hfill}%
\renewcommand{\@evenfoot}{\hfill\small{--~\thepage~--}\hfill}%
\renewcommand{\@oddfoot}{\hfill\small{--~\thepage~--}\hfill}%
}\makeatother\pagestyle{plain}


\title{~\vspace*{-12pt}\\\hspace*{-1pt}\mbox{Magnetic field dependence of the neutron spin resonance in CeB$_{6}$}}

\author{P.\hspace{0.6ex}Y.\hspace{0.6ex}Portnichenko}
\affiliation{Institut für Festkörperphysik, TU Dresden, D-01069 Dresden, Germany}

\author{S.\,V.~Demishev}
\affiliation{A. M. Prokhorov General Physics Institute of RAS, 38 Vavilov Street, 119991 Moscow, Russia}

\author{A.\,V.~Semeno}
\affiliation{A. M. Prokhorov General Physics Institute of RAS, 38 Vavilov Street, 119991 Moscow, Russia}

\author{H.~Ohta}
\affiliation{Department of Physics, Kobe University, Nada, Kobe 657-8501, Japan}

\author{A.~S.~Cameron}
\affiliation{Institut für Festkörperphysik, TU Dresden, D-01069 Dresden, Germany}

\author{M.~A.~Surmach}
\affiliation{Institut für Festkörperphysik, TU Dresden, D-01069 Dresden, Germany}

\author{H.~Jang}
\affiliation{Max-Planck-Institut für Festkörperforschung, Heisenbergstraße 1, D-70569 Stuttgart, Germany}
\affiliation{Stanford Synchrotron Radiation Lightsource, SLAC National Accelerator Laboratory, Menlo Park, California 94025, USA}

\author{G.~Friemel}
\affiliation{Max-Planck-Institut für Festkörperforschung, Heisenbergstraße 1, D-70569 Stuttgart, Germany}

\author{A.\,V.~Dukhnenko}
\affiliation{I. M. Frantsevich Institute for Problems of Material Sciences of NAS, 3 Krzhyzhanovsky Street, 03680 Kiev, Ukraine}

\author{N.~Yu.~Shitsevalova}
\affiliation{I. M. Frantsevich Institute for Problems of Material Sciences of NAS, 3 Krzhyzhanovsky Street, 03680 Kiev, Ukraine}

\author{V.~B.~Filipov}
\affiliation{I. M. Frantsevich Institute for Problems of Material Sciences of NAS, 3 Krzhyzhanovsky Street, 03680 Kiev, Ukraine}

\author{A.~Schneidewind}
\affiliation{J\"ulich Center for Neutron Science (JCNS), Forschungszentrum J\"ulich GmbH, Outstation at Heinz Maier\,--\,Leibnitz Zentrum~(MLZ), Lichtenbergstra{\ss}e 1, D-85747 Garching, Germany}

\author{J.~Ollivier}
\affiliation{Institut Laue-Langevin, 6 rue Jules Horowitz, BP 156, F-38042 Grenoble Cedex, France}

\author{A.~Podlesnyak}
\affiliation{Quantum Condensed Matter Division, Oak Ridge National Laboratory (ORNL), Oak Ridge, TN 37831-6475, USA}

\author{D.~S.~Inosov}\email[Corresponding author:\vspace{6pt}]{Dmytro.Inosov@tu-dresden.de}
\affiliation{Institut für Festkörperphysik, TU Dresden, D-01069 Dresden, Germany}

\begin{abstract}\parfillskip=0pt\relax
\noindent{In zero magnetic field, the famous neutron spin resonance in the $f$\!-electron superconductor CeCoIn$_{5}$ is similar to the recently discovered exciton peak in the nonsuperconducting CeB$_{6}$. A magnetic field splits the resonance in CeCoIn$_{5}$ into two components, indicating that it is a doublet. Here we employ inelastic neutron scattering (INS) to scrutinize the field dependence of spin fluctuations in CeB$_{6}$. The exciton shows a markedly different behavior without any field splitting. Instead, we observe a second field-induced magnon whose energy increases with field. At the ferromagnetic zone center, however, we find only a single mode with a nonmonotonic field dependence. At low fields, it is initially suppressed to zero together with the antiferromagnetic order parameter, but then reappears at higher fields inside the hidden-order phase, following the energy of an electron spin resonance (ESR). This is a unique example of a ferromagnetic resonance in a heavy-fermion metal seen by both ESR and INS consistently over a broad range of magnetic fields.}
\end{abstract}

\maketitle

\section{Introduction}\vspace{-5pt}

The observation of neutron spin resonance within a broad range of materials, in particular high-$T_{\rm c}$ cuprates \cite{Eschrig06}, iron pnictides \cite{Dai15, Inosov16}, and heavy-fermion superconductors \cite{SatoAso01, StockBroholm08, StockertArndt11}, is recognized as an indicator of unconventional superconductivity. It was shown that sign-changing gap symmetry can lead to the existence of resonance behavior \cite{BulutScalapino96, EreminZwicknagl08, MazinSingh08, HirschfeldKorshunov11}. Of particular interest are inelastic neutron scattering (INS) results obtained on CeCoIn$_{5}$, where a sharp resonance peak was observed within the superconducting phase \cite{StockBroholm08, StockBroholm12a, RaymondKaneko12, RaymondLapertot15}. At first glance similar peaks were found in the antiferromagnetic (AFM) superconductor UPd$_{2}$Al$_{3}$ \cite{BlackburnHiess06, HiessBernhoeft06}, as well as in the normal state of the heavy-fermion metal YbRh$_{2}$Si$_{2}$ \cite{StockBroholm12}, where superconductivity was recently discovered below $\sim2$\,mK \cite{SchuberthTippmann16}. Another striking example of a resonant mode is given by the well known nonsuperconducting heavy-fermion antiferromagnet CeB$_{6}$ \cite{FriemelLi12,JangFriemel14}. The microscopic origins of such resonant magnetic excitations persisting in $f$\!-electron systems either with or without superconductivity may well differ among materials and are still hotly debated.

The application of an external magnetic field may help to unmask the differences between these various excitations. For instance, among $f\!$-electron compounds, a weak quasielastic signal gives rise to a field-induced ferromagnetic (FM) excitation in CeRu$_{2}$Si$_{2}$ \cite{SatoKoike04}. In YbRh$_{2}$Si$_{2}$, two incommensurate excitation branches merge into a commensurate FM resonance whose energy scales linearly with magnetic field \cite{StockBroholm12}, whereas in UPd$_{2}$Al$_{3}$ the energy gap initially remains almost constant inside the superconducting phase, but starts following a monotonic linear dependence at higher magnetic fields \cite{BlackburnHiess06}. The sharp resonance in CeCoIn$_{5}$ splits into a Zeeman doublet \cite{StockBroholm12a} rather than a theoretically predicted triplet~\cite{AkbariThalmeier12}, whereas in Ce$_{1-x}$La$_x$B$_6$ the magnetic field reportedly leads to a crossover from an itinerant to a more localized behavior of spin fluctuations \cite{FriemelJang15}. Thus, the application of an external magnetic field is an important tool to distinguish different types of collective spin excitations and to develop microscopic theoretical models for the formation of resonant modes.

Apart from neutron spectroscopy, a complementary way of probing spin dynamics is electron spin resonance (ESR). For a long time it was believed that due to the effect of Kondo screening, no ESR signal could be observed in Kondo lattices, as spin-orbit coupling significantly shortens electron spin relaxation times, leading to a broad and weak signal \cite{SichelschmidtIvanshin03}. This established opinion was impugned when for the first time Yb$^{3+}$ resonance was observed in a dense Kondo lattice system \cite{SichelschmidtIvanshin03, SichelschmidtWykhoff07, SchaufusKataev09}. Various theoretical explanations proposed complementary models which explained the existence of the narrow ESR line \cite{AbrahamsWoelfle08, Schlottmann09, ZvyaginKataev09}, while further investigation of different Kondo lattice systems demonstrated that FM correlations are of principal importance for the observation of the ESR signal \cite{KrellnerFoerster08}. While this empirical result summarized observations from a limited number of $f$\!-electron compounds, no clear counterexamples are known to date.

Cerium hexaboride is a nonsuperconducting heavy-fermion metal with a simple-cubic crystal structure \cite{CameronFriemel16}. Competition between the Kondo screening and the Ruderman-Kittel-Kasuya-Yosida (RKKY) coupling mechanism via conduction electrons leads to a rich magnetic-field\,--\,temperature phase diagram. Its ground-state phase below $T_{\rm N}=2.3$\,K \cite{ZaharkoFischer03} is antiferromagnetic with a double-$\mathbf{q}$ structure, known as phase III, which undergoes a transition to single-$\mathbf{q}$ phase III$^\prime$ with the application of a magnetic field \cite{EffantinRossat-Mignod85}. Another phase transition at $T_{\rm Q}=3.2$\,K \cite{FujitaSuzuki80} corresponds to antiferroquadrupolar (AFQ) ordering in this compound (phase II), which was observed directly with resonant x-ray scattering \cite{NakaoMagishi01} as well as with neutron diffraction in a magnetic field \cite{EffantinRossat-Mignod85}. As both order parameters are presumably driven by AFM interactions, the observation of a sharp ESR signal within the AFQ phase came as a surprise and was then explained with ferromagnetically interacting localized magnetic moments \cite{DemishevSemeno06, DemishevSemeno09, DemishevSemeno08, Schlottmann12, Schlottmann13}. Only recently, an INS study revealed a strong FM mode in the magnetic excitation spectrum of CeB$_6$ \cite{JangFriemel14}, yet these observations were done in the zero-field AFM state, whereas ESR measurements could be only performed at elevated magnetic fields within the AFQ phase, precluding a direct comparison. In addition, a sharp resonant mode similar to that of CeCoIn$_5$ was revealed below $T_{\rm N}$ at the propagation wave vector of the AFQ phase \cite{FriemelLi12}, motivating a theoretical suggestion that the spin excitation spectrum of CeB$_6$ is dominated by the response of itinerant heavy quasiparticles rather than localized moments \cite{AkbariThalmeier12a, KoitzschHeming16}. However, a crossover to the localized-moment description was suggested for higher magnetic fields \cite{FriemelJang15}. Here we follow in detail the magnetic field dependence of spin excitations in CeB$_6$, including both FM and AFM spin resonances, across the quantum critical point (QCP) that separates the AFM and AFQ phases with the application of a magnetic field. Thus, our present observations bridge the gap between previous zero-field INS and high-field ESR measurements and provide a consistent description of spin dynamics that is clearly distinct from that known for other $f$\!-electron systems.

\begin{figure}[t]\vspace{-1.0em}
\includegraphics[width=0.9\columnwidth]{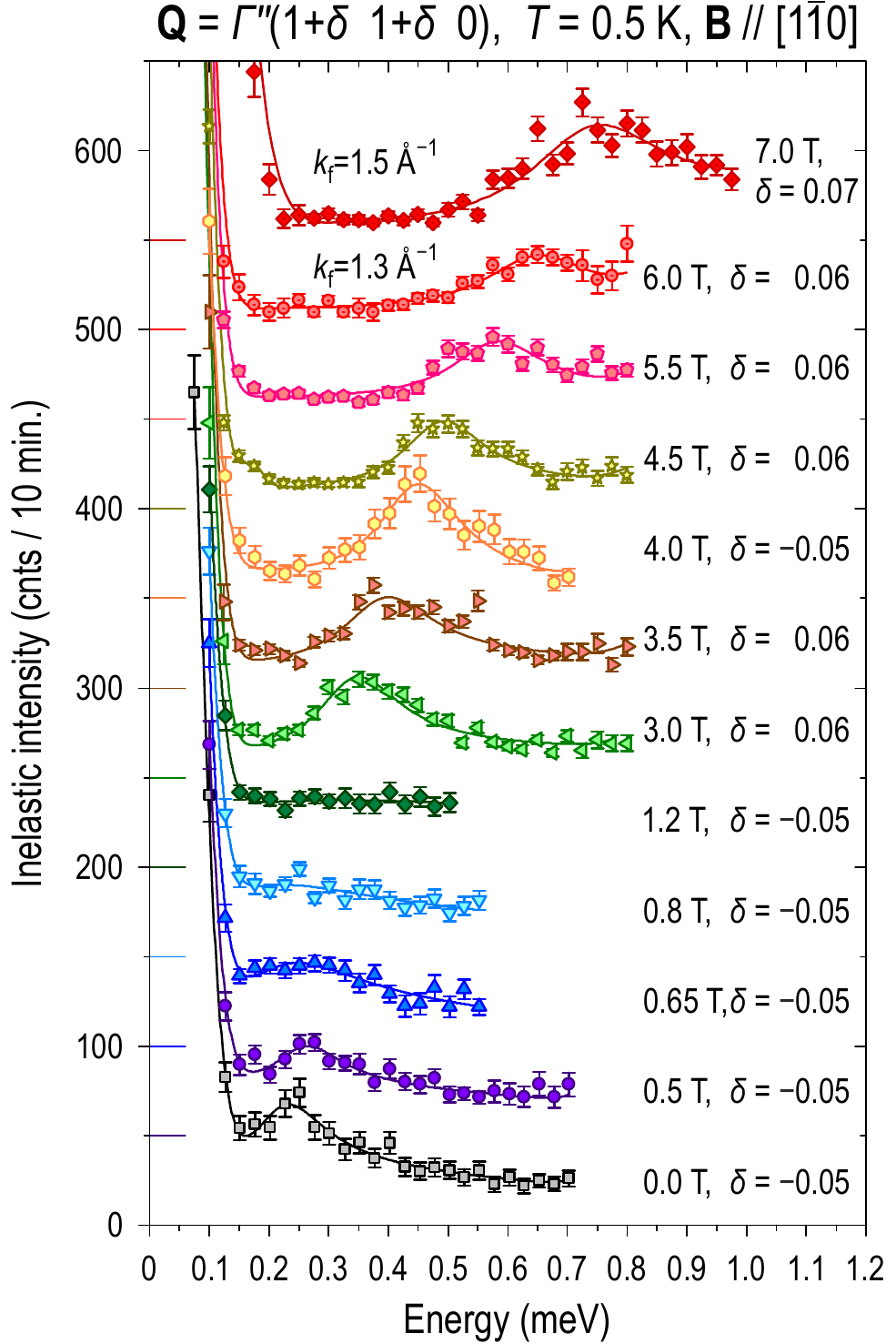}\vspace{-5pt}
\caption{INS spectra measured near the zone center $\Gamma''(110)$ at a slightly incommensurate wave vector as indicated in the legend, to avoid contamination from the Bragg tail. The spectra are shifted vertically for clarity with horizontal lines at the left indicating the background baseline for each spectrum. Solid lines represent Lorentzian fits on top of a nonmagnetic background.}
\label{Fig:Gamma}\vspace{-4pt}
\end{figure}

\vspace{-5pt}\section{Experimental results}\vspace{-5pt}
INS experiments were performed at the cold-neutron triple-axis spectrometer (TAS) PANDA \cite{PANDA} operated by JCNS at MLZ, Garching, the disk chopper time-of-flight (TOF) spectrometer IN5 \cite{OllivierMutk11} at ILL, Grenoble, and the cold-neutron chopper spectrometer (CNCS) \cite{EhlersPodlesnyak11} at the Spallation Neutron Source, ORNL. A rod-shaped single crystal of CeB$_{6}$ with a mass of 4\,g was grown by the floating-zone method from a 99.6\,\% isotope-enriched $^{11}$B powder (to minimize neutron absorption), as described elsewhere \cite{FriemelLi12}. We fixed the final wave vector of the neutrons to $k_{\rm f}=1.3$ or $1.5$\,\AA$^{-1}$ and used a cold Be filter to avoid higher-order neutron contamination for TAS experiments. TOF measurements were done with the incident neutron wavelength fixed at 5\,\AA\:(3.27\,meV) for IN5 and at 5.1\,\AA\:(3.15\,meV) for CNCS experiments. The sample environment comprised a 7.5\,T vertical-field cryomagnet with a $^{3}$He insert, 2.5\,T ``orange'' cryostat based magnet, and 5\,T cryomagnet for PANDA, IN5, and CNCS experiments, respectively. The ESR experiments were performed on a cavity spectrometer providing frequency range of 60--100\,GHz and magnetic field up to 7\,T (GPI, Moscow). Experiments at higher frequencies 100--360\,GHz using a 30\,T pulsed magnet were carried out at Kobe University with a quasioptical setup operating in reflection mode \cite{NakagawaYamada98}. The magnetic field for all experiments was aligned along the $[1\bar{1}0]$ direction of the crystal.

\begin{figure*}[t]\vspace{-1.2em}
\includegraphics[width=\textwidth]{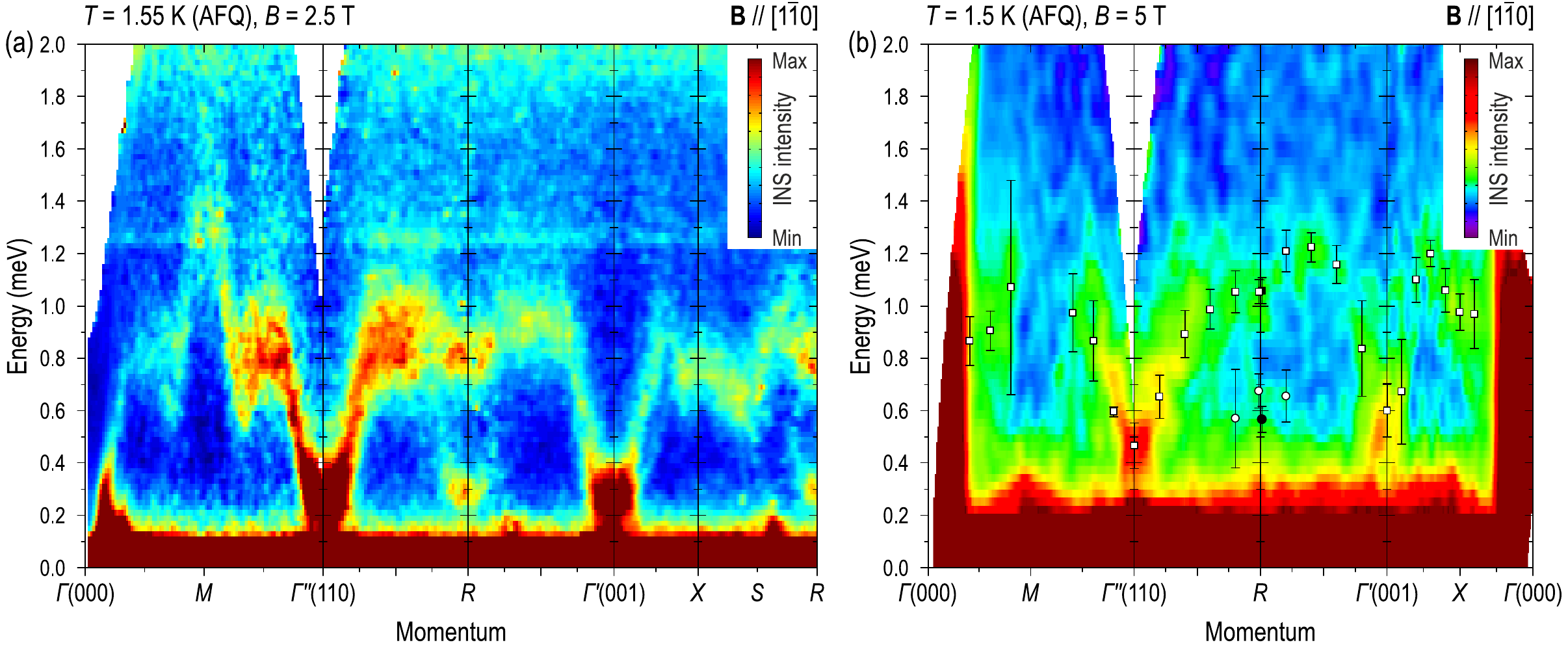}\vspace{-5pt}
\caption{Energy-momentum profiles along high-symmetry directions in the AFQ state: (a) $B = 2.5$\,T, (b) $B = 5$\,T. Open markers are determined as peak maxima from the fits. Solid markers at the $R(\frac{1}{2}\frac{1}{2}\frac{1}{2})$ point were obtained from the interpolation of peak positions from Ref.~\citenum{FriemelJang15}. Background contamination from the He exchange gas was subtracted from the data in panel (b), as explained in the Supplemental Material \cite{SupplementalMaterial16}. Because of the high-level background coming from the magnet, the field-induced low-energy magnetic excitation cannot be clearly resolved at this field.}
\label{Fig:QEmaps}\vspace{-5pt}
\end{figure*}

We first present the evolution of the FM resonance measured by TAS in magnetic fields up to 7\,T at $T=0.5$\,K. Figure~\ref{Fig:Gamma} shows unprocessed energy scans near the zone center $\Gamma''(1\!+\!\delta \;1\!+\!\delta\; 0)$. Slightly incommensurate wave vectors were chosen to avoid the contamination from phonons and the Bragg tail. The previously reported sharp resonance gets initially suppressed and broadens with the application of an external magnetic field as long as the system remains in the AFM state. The observed signal can be described by a Lorentzian line shape \cite{GoremychkinOsborn00}
\begin{multline}\label{Eq:Quasielastic}
S(\mathbf{Q},\omega)\propto\,\frac{\omega}{1-\exp(-\hslash\omega/k_\text{B}T)}\\
\times\biggl(\frac{\Gamma}{\hslash^2(\omega-\omega_0)^2+\Gamma^2}+\frac{\Gamma}{\hslash^2(\omega+\omega_0)^2+\Gamma^2}\biggr),
\end{multline}
where $\Gamma$ is the half width at half maximum of the Lorentzians centered at $\pm\,\hslash\omega_0$, whereas $\hslash$ and $k_{\rm B}$ are fundamental constants. Upon entering the phase III$^{\prime}$ at $\sim$\,1.2\,T~\cite{KunimoriKotani11}, the resonance is fully suppressed in energy and becomes quasielastic with $\hslash\omega_0=0$. However, the excitation reappears at higher magnetic fields within the AFQ phase at an energy that continuously increases with the applied field.

\begin{figure}[b]
\includegraphics[width=0.95\columnwidth]{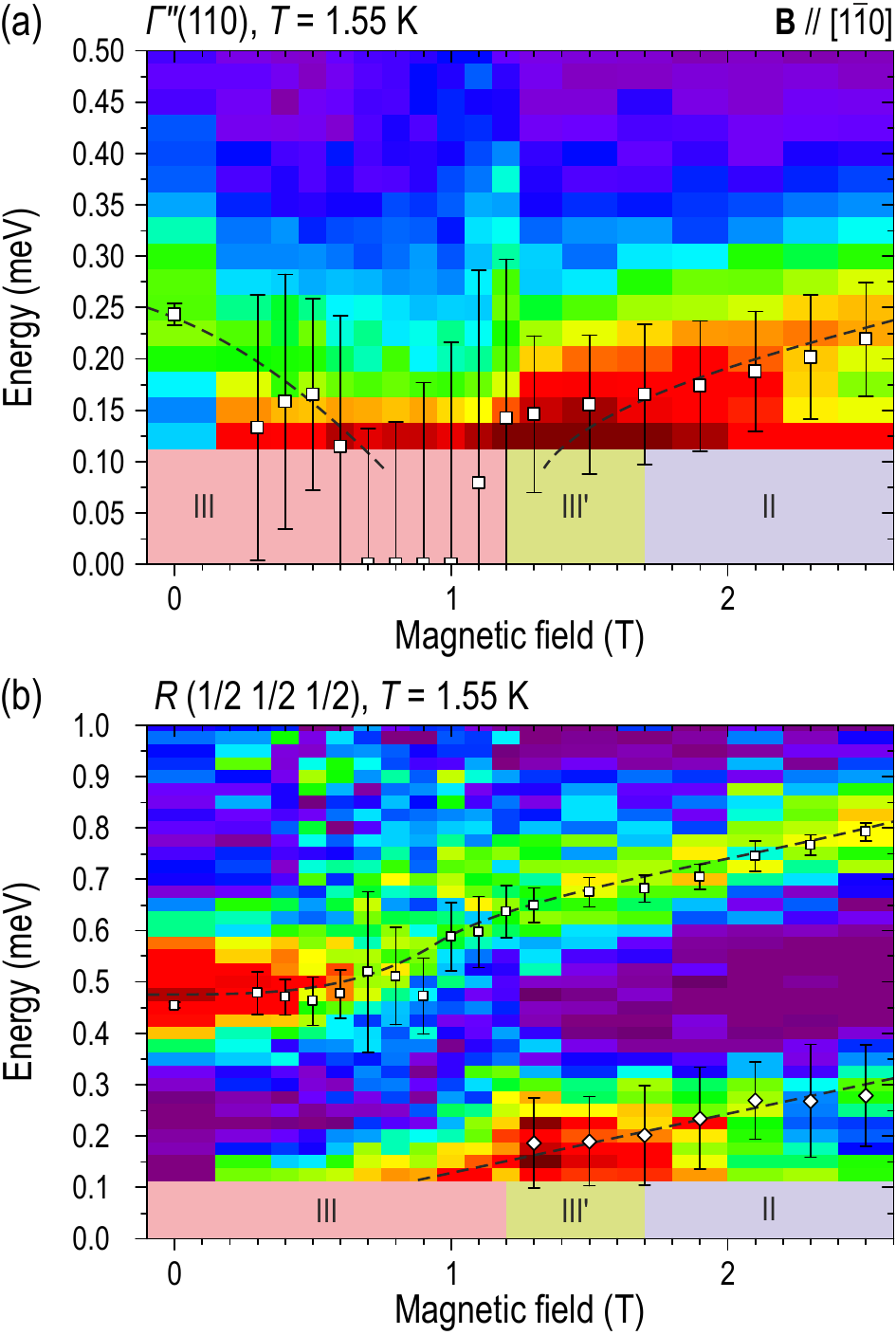}
\caption{Magnetic field dependence of the resonance peaks at (a) $\Gamma$ and (b) $R$ points. Markers in both panels were determined as peak maxima from the fits. Dashed lines are guides to the eyes, and the shaded areas below each panel mark the resolution cutoff and indicate the field regions corresponding to the AFM (III, III$^\prime$) and AFQ (II) phases.}
\label{Fig:Bscan}\vspace{-4pt}
\end{figure}

To get a more complete picture about the field dependence of magnetic excitations, we also performed TOF measurements on the same sample. A continuous dispersive magnon band connecting the local intensity maxima at the zone center ($\Gamma$) and zone corner ($R$) was observed at 2.5 and 5\,T, and its intensity distribution along main high-symmetry directions of reciprocal space is illustrated in Fig.~\ref{Fig:QEmaps}. It is remarkable that the magnon is more intense around the $\Gamma''(110)$ point than at the equivalent $\Gamma'(001)$ or $\Gamma(000)$ positions \cite{NotePrimes}, suggesting an anomalous nonmonotonic behavior of the dynamic form factor that is characteristic of multipolar moments (for conventional dipolar moments, it would decrease monotonically with $|\mathbf{Q}|$) \cite{KuramotoKusunose09, Shiina12, KuwaharaIwasa07}. A magnetic field of 2.5\,T [Fig.~\ref{Fig:QEmaps}(a)] does not change the excitation energy at the zone center significantly but increases the magnon bandwidth twofold, as the dispersion now reaches $\sim$\,1.4\,meV at the $M$ point in contrast to 0.7\,meV in zero field \cite{JangFriemel14}. In addition, a second field-induced low-energy magnetic excitation appears at the AFQ propagation vector, $R\,(\frac{1}{2}\frac{1}{2}\frac{1}{2})$. At first glance, the two modes at the $R$ point, separated by 0.5~meV, are reminiscent of the resonance-peak splitting in CeCoIn$_5$, yet our discussion hereinafter will demonstrate that the origin of this splitting is qualitatively distinct. An even higher magnetic field of 5\,T [Fig.~\ref{Fig:QEmaps}(b)] leads to a nearly twofold increase of the zone-center spin gap. We also note that the clear local maximum of intensity at the $R$ point \cite{FriemelLi12, JangFriemel14} is no longer seen at this field, indicating that the resonant exciton mode is suppressed and becomes part of the more conventional \mbox{magnon spectrum emanating from the zone center}.

Evolution of the magnetic excitations at the $\Gamma$ and $R$ points as a function of field would complement our TAS data and reveal essential differences in the behavior of the resonances in comparison with other heavy-fermion systems. Hence we focused our attention mainly on the $\Gamma$ and $R$ points and measured in detail the field dependence across the QCP using the TOF spectrometer IN5 equipped with a low-background 2.5 T cryomagnet. Energy-momentum profiles for each field along the $\Gamma R$ direction are shown as an animation in the Supplemental Material \cite{SupplementalMaterial16}. We also present one-dimensional energy profiles obtained from the same data by integration within $\pm\,0.15$~r.l.u. around the $\Gamma$ and $R$ points as color maps in Fig.~\ref{Fig:Bscan}. The data in Fig.~\ref{Fig:Bscan}(a) illustrate the nonmonotonic behavior of the zone-center excitation as it initially softens to zero upon entering the phase III$^\prime$ and then reappears within phase II at an energy that continuously increases with the applied field. A qualitatively different picture is observed for the resonance peak at the $R$ point in Fig.~\ref{Fig:Bscan}(b). Increasing the field within phase III keeps the resonance energy constant while it decreases in amplitude and broadens, transferring a significant part of its spectral weight to the second low-energy mode whose tail can be seen above the elastic line already above $\sim$\,0.5\,T. Upon crossing through the phase III--III$^\prime$ transition, the amplitude of the low-energy mode is maximized, whereas the higher-energy mode shifts up in energy. Both excitations then follow a linear trend with the same slope and approximately equal amplitudes in phase II, in agreement with our earlier report \cite{FriemelJang15}. This behavior is completely different from the field-induced splitting of the neutron resonance in the SC state of CeCoIn$_{5}$, where the second mode emerges from the resonance energy and then shifts down monotonically with increasing field \cite{StockBroholm12a}.

\begin{figure}[t]\vspace{-6pt}
\includegraphics[width=\columnwidth]{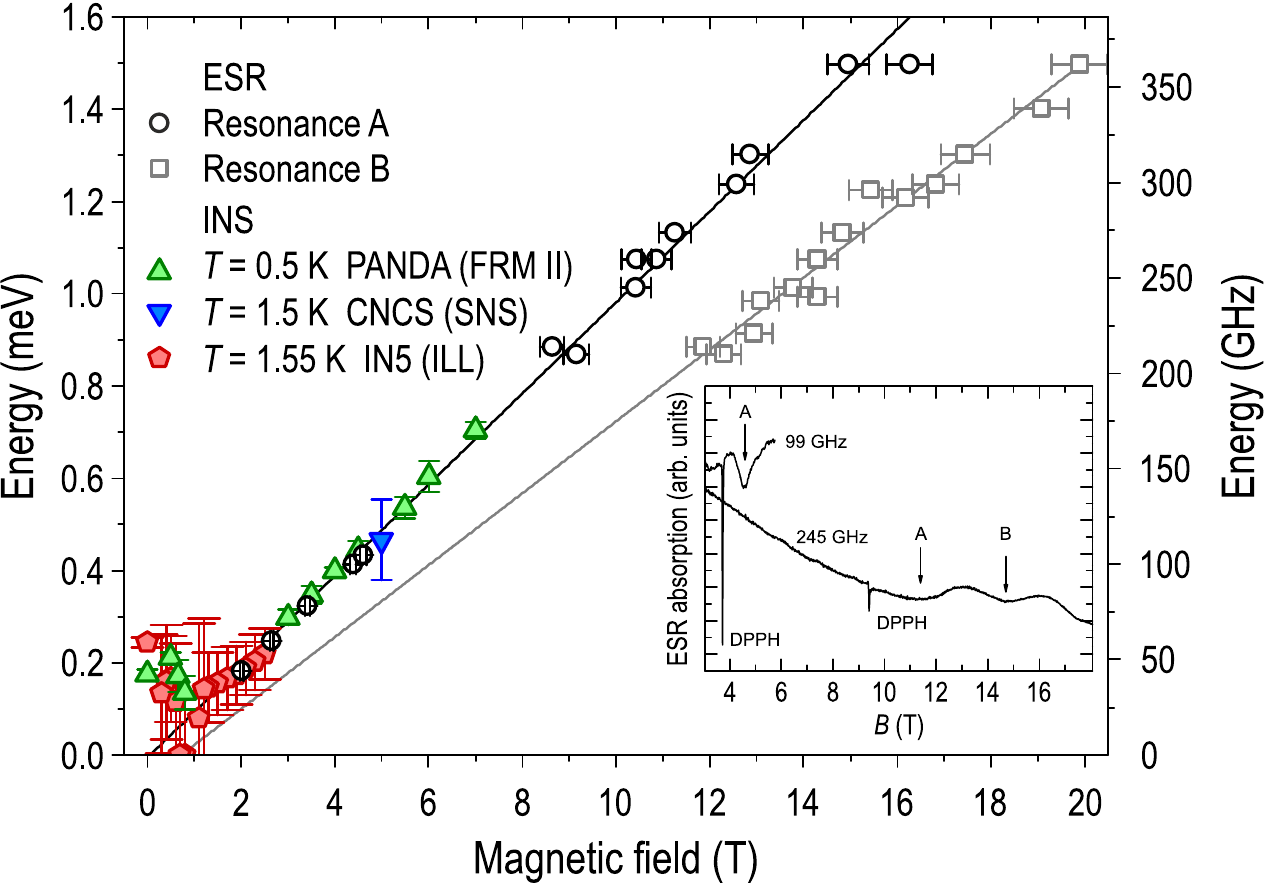}\vspace{-3pt}
\caption{Summary of the magnetic field dependence of zone-center excitations obtained from both INS and ESR spectra. Solid lines are linear fits of resonances A and B. The inset shows a field dependence of the cavity transmission at 99\,GHz and ESR spectrum obtained at 245\,GHz using a quasioptical technique as typical examples of unprocessed data sets from which the points in the main plot were obtained. Sharp lines marked as DPPH originate from a small 2,2-diphenyl-1-picrylhydrazyl reference sample.}
\label{Fig:ESR}
\end{figure}

ESR measurements, which probe zone-center excitations, have shown that the frequencies of the two observed resonances A and B \cite{DemishevSemeno08} change linearly with field within phase~II, as shown in Fig.~\ref{Fig:ESR} with open symbols. The linear fits shown with solid lines, $\hslash\omega\kern-.2pt=\kern-.2pt\hslash\omega_{0}\kern-.3pt+{\kern-.3pt}g\mu_{\rm B}B$, result in slopes of 0.098(2) and 0.077(4)\,meV/T for the resonances A and B \cite{DemishevSemeno08}, corresponding to $g$ factors of 1.70(4) and 1.35(7), respectively, as compared to that of 1.90(7) at the $R$ point \cite{FriemelJang15}. In Fig.~\ref{Fig:ESR} we compare resonance energies obtained from ESR (open symbols) with the field-dependent energy of the zone-center INS excitation (solid symbols). We find perfect agreement between the INS data and the resonance A in the intermediate field range within phase II, where both data sets overlap, suggesting that the same FM excitation is probed in both experiments. This comparison nicely demonstrates the complementarity of the ESR and INS methods.

\vspace{-10pt}\section{Conclusions}\vspace{-5pt}
In summary, we have investigated magnetic field dependencies of collective magnetic excitations at the zone center ($\Gamma$) and zone corner ($R$), as well as the ESR signal in CeB$_{6}$. Unlike in CeCoIn$_{5}$, where the AFM resonance splits into a Zeeman doublet, in CeB$_{6}$ the second field-induced magnon at the $R$ point exhibits a monotonically increasing field dependence. The FM resonance at the $\Gamma$ point is initially suppressed in energy with the magnetic field within the AFM phase, but reappears upon entering the AFQ phase. Its energy matches that of the resonance A seen in ESR, whereas the anomalous dynamic form factor of the zone-center excitation points towards its multipolar-wave character. This observation is consistent with the proposed orbital-ordering nature of the ESR response, resulting from the interplay of AFQ order with FM correlations \cite{DemishevSemeno08, Schlottmann12, Schlottmann13}, and with the multipolar character of phase II. The second ESR line observed in high fields (mode B) was interpreted as the result of a crossover of the excited state to the free-ion limit, as the field at which it appears is comparable with the condensation energy of the AFQ phase, $\sim\!1.75{\kern.5pt}k_{\rm B}T_{\rm Q}$ \cite{Schlottmann13}. The field available in our INS measurements was so far insufficient to reach this regime, therefore it still remains an open question if this second resonance might also appear in the INS spectra above 12\,T. Our current results are a rare example of a simultaneous observation of the FM resonance by INS and ESR, consistently over a broad range of magnetic fields, thus demonstrating the complementarity of these two spectroscopic probes.

\vspace{-5pt}\section{Acknowledgments}\vspace{-5pt}
We acknowledge stimulating discussions with V.~Kataev and thank S.~Elorfi (SNS) for technical support during the experiments. TOF data reduction was done using the \emph{Horace} software package \cite{Horace}. This project was funded by the German Research Foundation (DFG) under grant No.~IN\,\mbox{209/3-1} and via the Research Training Group GRK\,1621 at the TU Dresden. S.\,V.~D. and A.\,V.~S. acknowledge support from the RFBR grant 14-02-00800 and from the RAS Programmes ``Electron spin resonance, spin-dependent electronic effects and spin technologies'' and ``Electron correlations in strongly interacting systems''. Research at the Spallation Neutron Source in Oak Ridge was supported by the Scientific User Facilities Division, Office of Basic Energy Sciences, US Department of Energy.\vspace{-6pt}

\bibliographystyle{my-apsrev}\bibliography{references}

\begin{thebibliography}{55}
\bibitem [{Not()}]{NotePrimes}%
  \BibitemOpen
  \bibinfo {howpublished} {We label all high-symmetry points of the Brillouin
  zone according to the standard notation for the cubic system. In consistency
  with previous works \cite{FriemelLi12, JangFriemel14, PortnichenkoCameron15,
  PortnichenkoRomhanyi16}, we add primes to distinguish between equivalent wave
  vectors in the first, second, and third Brillouin zones that are
  characterized by different lengths of the scattering vector and, hence, have
  different form factors. Here, for example, $\Gamma(000)$ denotes the center
  of the first Brillouin zone, $\Gamma'(001)$ is the nearest equivalent
  reciprocal-space vector at the center of the second Brillouin zone,
  $\Gamma''(110)$ is the next-nearest equivalent point, and so on.}\BibitemShut
  {Stop}%
\end{thebibliography}

\onecolumngrid
\clearpage

\pagestyle{plain}

\renewcommand\thefigure{S\arabic{figure}}
\renewcommand\thetable{S\arabic{table}}
\renewcommand\theequation{S\arabic{equation}}
\renewcommand\bibsection{\section*{\sffamily\bfseries\footnotesize Supplementary References\vspace{-6pt}\hfill~}}

\citestyle{supplement}

\makeatletter\immediate\write\@auxout{\string\bibstyle{my-apsrev}}\makeatother

\pagestyle{plain}
\makeatletter
\renewcommand{\@oddfoot}{\hfill\bf\scriptsize\textsf{S\thepage}}
\renewcommand{\@evenfoot}{\bf\scriptsize\textsf{S\thepage}\hfill}
\renewcommand{\@oddhead}{P.\hspace{0.6ex}Y.\hspace{0.6ex}Portnichenko \textit{et~al.}\hfill\Large\textsf{{Supplemental Material}}}
\renewcommand{\@evenhead}{P.\hspace{0.6ex}Y.\hspace{0.6ex}Portnichenko \textit{et~al.}\Large\textsf{{Supplemental Material}}\hfill}
\renewcommand{\fnum@figure}[1]{\figurename~\thefigure.}

\makeatother
\setcounter{page}{1}\setcounter{figure}{0}\setcounter{table}{0}\setcounter{equation}{0}

\onecolumngrid\normalsize
\begin{center}{\vspace*{0.1pt}\Large{Supplemental Material to the Letter\smallskip\\\sl\textbf{``\hspace{1pt}Magnetic field dependence of the neutron spin resonance in CeB$_{6}$''}}}\end{center}\bigskip

\twocolumngrid

\begin{figure}[t]\vspace{-5pt}
\noindent\animategraphics[controls, loop, width=\columnwidth, timeline=timeline.txt]{2}{figure_S}{1}{18}
\caption{Animation illustrating magnetic field dependence of the magnon spectrum along high-symmetry directions in CeB$_6$. Each frame consists of three color maps integrated along straight segments connecting the $X(00\frac{1}{2})$, $R(\frac{1}{2}\frac{1}{2}\frac{1}{2})$, $\Gamma''(110)$, and $X''(11\frac{1}{2})$ points, which we combined to form a continuous polygonal path in reciprocal space. The controls at the bottom of the figure can be used to modify the frame rate, pause the animation, or browse through individual frames.}
\label{Fig:Animation}\vspace{-0.3em}
\end{figure}

\bigskip
\noindent\textbf{\it Detailed magnetic field dependence}\vspace*{5pt}

Here in Fig.~S1 we present an animation showing the full set of INS data on CeB$_6$ measured at the cold-neutron TOF spectrometer IN5 as a function of magnetic field. These data were obtained as two-dimensional cuts along the $(H\,H\,\frac{1}{2})$, $(H\,H\,1\!-\!H)$, and $(1\,1\,L)$ high-symmetry directions from our four-dimensional TOF data set by integrating within $\pm$\,0.15~r.l.u. along the momentum direction perpendicular to the plane of the figure in the $(HHL)$ scattering plane, and $\pm$\,0.04~r.l.u. in the out-of-plane (vertical) direction parallel to the magnet axis. Evolution of both resonant modes at the zone center ($\Gamma$) and zone corner ($R$) can be observed.

\begin{figure}[b]\vspace{-5pt}
\includegraphics[width=\columnwidth]{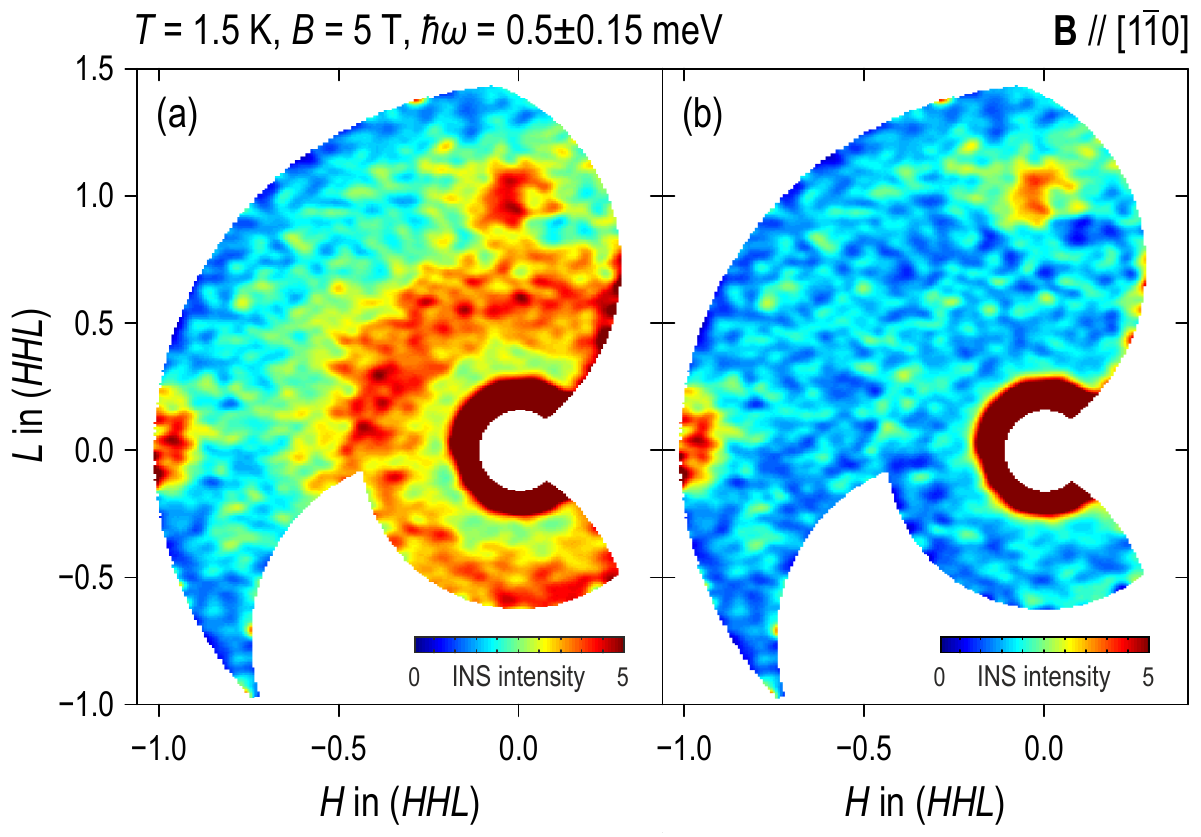}
\caption{Constant-energy map of the INS intensity, obtained by integrating the TOF data measured at CNCS in the energy window [0.35 0.65]\,meV. The data show resonant FM excitations centered at $\Gamma''(110)$ and $\Gamma'(001)$: (a) as initially measured, (b) after substraction of the He background according to Eq.\,(\ref{Eq:He_gas}).}
\label{Fig:He_sub}\vspace{-4pt}
\end{figure}

\bigskip
\noindent\textbf{\it Comments on cryomagnet background.}\vspace*{5pt}

The signal in INS experiments can be contaminated by artifacts originating from the sample environment, higher-order neutrons from the monochromator, or accidental Bragg scattering. Sample holders and cryogenic sample equipment are typically made of aluminum [\ref{ShiraneShapiro02}] due to its low absorption and incoherent scattering cross-section, which helps to minimize the background. Contamination from the Al powder lines that originates from the strongest $(1 1 1)$, $(2 0 0)$, $(2 2 0)$, and $(1 1 3)$ Bragg reflections appears at $|\mathbf{Q}|>2.69$\,\AA$^{-1}$, and due to the kinematic constraints is therefore not observed in our data. However, incoherent scattering of the incident neutron beam on the Al walls of a cryostat or cryomagnet can significantly broaden the elastic line, as it typically appears in the spectrum as a pair of peaks slightly shifted to positive and negative energy transfer values with respect to the elastic position. Their intensities depend strongly on the thickness of the inner walls of the cryogenic device, the purity of the used aluminum, and the degree of collimation of the incident neutron beam. This type of background contamination was observed both in the TAS and TOF data, and we had to include the corresponding peaks in the fitting model for the shape of the elastic line to accurately describe our experimental results. Moreover, in the TOF experimental geometry, the opposite segments of the cryostat's inner wall that are illuminated by the direct neutron beam are seen by the detector at an increasing angle from each other as one goes to higher scattering angles. As a result, the apparent shape of the elastic line also becomes momentum-dependent after the data are transformed into energy-momentum space. We can observe this effect in Fig.~\ref{Fig:QEmaps}, where thicker Al walls of the cryomagnet at CNCS in Fig.~\ref{Fig:QEmaps}(b) produce more background and broaden the elastic line significantly in comparison to the IN5 data in Fig.~\ref{Fig:QEmaps}(a) that were collected using a low-background 2.5\,T cryomagnet.

\bigskip
\noindent\textbf{\it He exchange gas background substraction.}\vspace*{5pt}

Additional background contamination may originate from the He gas used for heat exchange between the sample and the cryostat in some types of the cryogenic sample environment [\ref{ShiraneShapiro02}]. It originates from nonmagnetic scattering on free He nuclei and is both momentum and temperature dependent, which can lead to a misinterpretation of experimental results [\ref{BuyersKjems85_s},\,\ref{BuyersKjems86_s}]. This type of background contamination was observed in our TOF data because of the ambient-pressure He exchange gas in the CNCS cryomagnet. It appears as a ring of intensity within every constant-energy cut, as can be seen in Fig.\,S2(a), where the contamination is observed in the momentum range 0.25~r.l.u.~$\leq |\mathbf{Q}| \leq$~0.5~r.l.u. Scattering from single free nuclei can be analytically described as [\ref{Squiresbook}]:
\begin{multline}\label{Eq:He_gas}
S(\mathbf{Q},\omega)\propto\,\biggl(\frac{\beta}{4\pi E_{\rm r}}\biggl)^{1/2}
\times
\exp\biggl[-\frac{\beta}{4E_{\rm r}}(\hbar\omega-E_{\rm r})\biggr]
\end{multline}
where $\beta=(k_{\rm B}T)^{-1}$ and $E_{\rm r}=\hslash^2|\mathbf{Q}|^2/(2M)$ is the recoil energy, $M$ being the mass of the nuclei. The CNCS data presented in Fig.~\ref{Fig:QEmaps}(b) have been background-corrected by subtracting the analytical form of the He signal given by Eq.\,(\ref{Eq:He_gas}). The amplitude of this contribution was kept as a free parameter and adjusted to provide the best fit to the measured data. For the purpose of this fitting, we have restricted the TOF data set to the volume of energy-momentum space that contained no Bragg reflections and no magnetic signal. The quality of the resulting subtraction is demonstrated by Fig.~\ref{Fig:He_sub}, where we compare constant-energy cuts through the 5\,T data set, integrated around the energy transfer of 0.5\,meV, before and after the described background-correction procedure. This analysis reveals the FM resonances at the equivalent $\Gamma''(110)$ and $\Gamma'(001)$ points, as expected for this energy.\vspace{-0.9em}\vfill

\bibsection\small
\begin{enumerate}[{[S1]},ref={S\arabic*},leftmargin=20.5pt,labelsep=4pt,itemsep=-2pt]
\item\label{ShiraneShapiro02} G.~Shirane, S.~M.~Shapiro, and J.~M.~Tranquada \textit{Neutron scattering with a triple-axis spectrometer}, (Cambridge University Press, 2002), ISBN 0-521-41126-2.
\item\label{BuyersKjems85_s} W.~J.~L.~Buyers, J.~K.~Kjems, and J.~D.~Garrett, \href{http://link.aps.org/doi/10.1103/PhysRevLett.55.1223}{\prl {\bf 55}, 1223 (1985)}.
\item\label{BuyersKjems86_s} W.~J.~L.~Buyers, J.~K.~Kjems, and J.~D.~Garrett, \href{http://link.aps.org/doi/10.1103/PhysRevLett.56.996.2}{\prl {\bf 56}, 996 (1986)}.
\item\label{Squiresbook} G.~L.~Squires, \textit{Introduction to the Theory of Thermal Neutron Scattering}, (Dover Publications, Inc., Mineola, New York, 1996), ISBN 0-486-69447-X.
\end{enumerate}\vspace{-0.85em}
\onecolumngrid

\end{document}